\documentclass[]{aspm}

\articleinfo{}{}{}

\usepackage{verbatim}
\usepackage{amssymb}
\usepackage{amsbsy}
\usepackage{amscd}
\usepackage{amsmath}
\usepackage{amsthm}
\usepackage[mathscr]{eucal}

\numberwithin{defn}{section}

\def\be{\begin{equation}}
\def\ee{\end{equation}}
\def\ba{\begin{eqnarray}}
\def\ea{\end{eqnarray}}

\def\tg{{\tilde g}}

\title[Separability of Maxwell eq.\ in rotating BH spt.\ and its geometric aspects]{Separability of Maxwell equation in rotating black hole spacetime and its geometric aspects}

\author[T. Houri]{Tsuyoshi Houri}
\author[N. Tanahashi]{Norihiro Tanahashi}
\author[Y. Yasui]{Yukinori Yasui}

\address{National Institute of Technology, Maizuru College, Kyoto 625-8511, Japan}
\address{Institute of Mathematics for Industry, Kyushu University, Fukuoka 819-0395, Japan}
\address{Institute for Fundamental Sciences, Setsunan University, Osaka 572-8508, Japan}

\email{t.houri@maizuru-ct.ac.jp}
\email{tanahashi@imi.kyushu-u.ac.jp}
\email{yukinori.yasui@mpg.setsunan.ac.jp}

\rcvdate{***}
\rvsdate{***}



\begin{document}

\begin{abstract}
Recently a new formalism for perturbations of Maxwell's equations on the background of the Kerr-NUT-(A)dS spacetime was proposed, with which the equations are reduced to a equation of motion of a scalar field that can be solved by separation of variables.
In this formalism the differential operators that commute with the operators of the equations of motion, called symmetry operators, played a key role to establish the separable structure. 
In this work we propose
a method to reproduce these commuting symmetry operators in terms of the geometric quantities associated to the hidden symmetry of the background spacetime.
\end{abstract}

\maketitle

\section{Introduction}

Black holes are basic objects in the gravitational theory, and their various aspects have been studied intensively.
Particularly perturbative dynamics of fields in black hole spacetime has attracted much interest
since it governs astrophysical phenomena around black holes and also shows intriguing mathematical properties.
In this context, the separability of the perturbation equations plays a key role
since it reduces the perturbation equations given as partial differential equations to
to a set of ordinary differential equations that may be solved analytically.

In this work we focus on spacetimes with rotating black holes~\cite{Kerr:1963ud,Myers:1986un}, in which the perturbation equations become complicated
due to the rotation.
For the Kerr black holes in four dimensions,
Teukolsky made a significant progress in his seminal work,
which showed that the equations of motion for the gravitational and Maxwell field can be separated
if they are expressed in the so-called Newman-Penrose formalism~\cite{PhysRevLett.29.1114,Teukolsky:1973ha}.
However, 
 this formalism works only in the four dimensions
and generalization to higher dimensions remained unaccomplished for a long time.

Recently a breakthrough about this problem was made by Lunin~\cite{Lunin:2017drx} by
introducing new variables to express the vector potential, which is the dynamical variables of the Maxwell field. In terms of these new variables, the perturbation equations can be separated for all the propagating modes, and also this formalism can be generalized to the higher-dimensional case straightforwardly.
%
%
In this work symmetries of the rotating black hole spacetime played an important role. It is known that the Kerr spacetime and its higher-dimensional generalization have hidden symmetries signified by the existence of nontrivial symmetric tensor satisfying
$\nabla_{(a}K_{bc)}=0$.
This tensor is called the Killing tensor, and it yields a constant of motion $K^{ab}p_a p_b$, where $p^a$ is the momentum vector, that is preserved on a geodesic. 
In Lunin's work, this tensor was used to introduce special coordinates that helped separating the variables of the perturbation equations.

Krtou\v{s}, Frolov and Kubiz\v{n}\'{a}k
made a step forward about clarifying the significance of the hidden symmetries~\cite{Krtous:2018bvk} in this problem.
They focused on the off-shell Kerr-NUT-AdS metric in $2n$ dimensions
\begin{equation}
g_{ab} dx^a dx^b=
  \sum_{\mu=1}^n \frac{U_\mu}{X_\mu} dx_\mu^2
+ \sum_{\mu=1}^n \frac{X_\mu}{U_\mu} \left(
\sum_{k=0}^{n-1} A^{(k)}_\mu d\psi_k
\right)^2 ~,
\label{Kerr-NUT-AdS}
\end{equation}
where $U_\mu$, $X_\mu$ and $A^{(k)}_\mu$ are certain functions of $x^\mu$.
This metric is the most general one admitting the principal tensor $h_{ab}$
which is an anti-symmetric generalization of the Killing tensor defined by
\begin{equation}
\nabla_c h_{ab} =
g_{ca} \xi_b - g_{cb} \xi_a ~,
\qquad
\xi_a = \frac{1}{2n-1} \nabla^b h_{ba} ~ ,
\end{equation}
where $\nabla_a$ is the covariant derivative (Levi-Civita connection) of $g_{ab}$.
Then they showed that, by expressing the vector potential expressed with a scalar $Z$
and a constant $\beta$
as
\begin{equation}
{\mathsf A}^a = B^{ab}\nabla_b Z ~,
\qquad
B^{ac}\left(g_{cb} - \beta h_{cb}\right) = \delta^a_b ~,
\end{equation}
Maxwell's equation in the Lorenz gauge
$\nabla_a {\mathsf A}^a = \nabla_a\left(B^{ab}\nabla_b Z\right)= 0$
can be reduced to a single scalar equation given by
\begin{equation}
\square Z + 2\beta \xi_c B^{cd}\nabla_d Z
= 0~,
\label{Maxwell}
\end{equation}
where $\square \equiv g^{ab}\nabla_a \nabla_b$ is the d'Alembertian operator.
In our work,
we call it the Lunin-Krtou\v{s}-Frolov-Kubiz\v{n}\'{a}k (LKFK) equation.
They also showed that these equations can be expressed using differential operators whose commutators vanish,
which makes it possible to solve the equations by separation of variables.
This result suggests that the separability of the perturbation equation and the hidden symmetry of the spacetime are intimately related with each other, although their precise correspondence still remains unclear.

In our work,
we shed new light on this problem based on
studies about the hidden symmetry of the spacetime and the separability.
One technique useful here is the Eisenhart-Duval lift~\cite{Eisenhart0,Duval:1984cj},
in which the original spacetime is supplemented with additional dimensions as
\begin{equation}
ds^2 = \tilde g_{AB}d x^A d x^B
= g_{ab}dx^a dx^b + 2 qA_a dx^a du + 2 du dv - 2 V du^2,
\label{lifted}
\end{equation}
where $g_{ab}$ is the original spacetime metric and $u, v$ are the coordinates for the additional spacetime dimensions.
By suitably choosing the metric functions $A^a, V$,
the derivative operator $\square  + 2\beta \xi_c B^{cd}\nabla_d $ in (\ref{Maxwell}) can be expressed simply as the d'Alembertian $\tilde \square = \tilde g^{AB} \tilde \nabla_A \tilde \nabla_B$ in terms of the covariant derivative $\tilde \nabla_A$ associated with $\tilde g_{AB}$.
Hence, with this technique we can reduce the perturbation equation~(\ref{Maxwell}) to a massless Klein-Gordon equation by absorbing the additional term $2\beta \xi_c B^{cd}\nabla_d $ into the structure of the higher-dimensional spacetime.

It turns out that Eq.~(\ref{lifted}) falls into the standard form of the metric whose geodesics is completely integrable~\cite{MR1156532}, hence it is associated with the Killing tensors $\tilde K_{AB}^{(i)}$ corresponding to the constants of motion on those geodesics.
Using these Killing tensors we can construct differential operators $\tilde {\mathcal K}^{(i)}\equiv \tilde \nabla_A \tilde K_{(i)}^{AB} \tilde\nabla_B$,
and it can be shown that these operators commute with the d'Alembertian 
$\bigl[\tilde \square,\tilde{\mathcal K}^{(i)}\bigr] = 0$
and also that they commutes with each other 
$\bigl[\tilde{\mathcal K}^{(i)},\tilde{\mathcal K}^{(j)}\bigr] = 0$.
It implies that 
$\tilde {\mathcal K}^{(i)}$
act as commuting symmetry operators that maps a solution of (\ref{Maxwell}) to another one, which plays a key role in separation of variables.
It can be further confirmed that 
the differential operators $\tilde {\mathcal K}^{(i)}$ coincide with
the commuting derivative operators found in \cite{Krtous:2018bvk}, where only the explicit form of the operators in terms of the coordinates were given.
Hence we found a way to give a geometric interpretation to the commuting symmetry operators found by \cite{Krtous:2018bvk}.

\section{Construction of commuting symmetry operators}
\label{sec:sec2}


\subsection{Eisenhart-Duval lift of a classical mechanical system}

In the Hamiltonian formulation, the motion of a relativistic particle with mass $m$ and charge $q$ in the presence of vector and scalar potentials $A_a$ and $V$ on a $D$-dimensional spacetime $(M,g_{ab})$, on which the line element is given by
 $ds^2 = g_{ab}dx^a dx^b $,
is governed by the Hamiltonian
\be
 H = \frac{1}{2}g^{ab}\left(p_a - qA_a\right)
 \left(p_b - qA_b\right) + V
 \label{originalHamiltonian}
\ee
with the constraint $H=-m^2$.
For such a system,
we may take the Eisenhart-Duval lift by considering
the $(D+2)$-dimensional manifold $\tilde{M} = M \times R^2$ with the metric%
\begin{align}
d\tilde{s}^2
= \tg_{AB}dx^Adx^B
= g_{ab}dx^a dx^b + 2 qA_a dx^a du + 2 du dv - 2 V du^2 \,,
\label{liftedmetric}
\end{align}
where
$ 
\tg_{ab} = g_{ab}(x^a) \,,
 \tg_{a u} = qA_a(x^a) \,, 
 \tg_{uv} = 1 \,, 
 \tg_{uu} = - 2V(x^a)$.
The geodesic Hamiltonian 
on the uplifted spacetime $(\tilde{M},\tg_{AB})$ is written as
\be
 \tilde{H} = \frac{1}{2}\tg^{AB}p_Ap_B
 = \frac{1}{2}g^{ab} (p_a-qA_a p_v)(p_b -qA_b p_v)
+ V p_v^2 + p_u p_v \,,
\label{liftedHamiltonian}
\ee
where
$ 
\tg^{ab} = g^{ab} \,, 
 \tg^{a v}= -q g^{ab}A_b \,, 
 \tg^{uv} = 1 \,, 
 \tg^{vv} = q^2A_a A^a + 2V $.
Since $g_{ab}$, $A_a$ and $V$ are functions of $x^a$ only,
$\partial/\partial u$ and $\partial/\partial v$ are Killing vector fields;
hence, the corresponding momenta $p_u$ and $p_v$ are constants.
Putting $\tilde{H}=0$, $p_u=m^2$ and $p_v=1$,
we obtain the equations of motion for the Hamiltonian (\ref{originalHamiltonian})
satisfying $H = -m^2$
from (\ref{liftedHamiltonian}). Thus null geodesics on the uplifted spacetime ($\tilde{M},\tg_{AB}$) project onto the solutions of the original system in $M$.

\subsection{Application to quantum mechanical systems}
In quantum mechanics, via the Dirac quantization $p_a \to -i\nabla_a$,
the equation of motion
corresponding
to the system of
the Hamiltonian (\ref{originalHamiltonian}) is obtained as%
\be
 {\mathcal H}\psi \equiv \Big[g^{ab}(\nabla_a - iqA_a)
 (\nabla_b - iqA_b) -2 V\Big]\psi =
-m^2
\psi
\,,
 \label{originalSchEq}
\ee
where $\nabla_a$ is the Levi-Civita connection on $(M,g_{ab})$.
A remarkable fact is that solutions to the equation of motion
 (\ref{originalSchEq})
are reproduced from a particular class of
the solutions to the massless Klein-Gordon equation on the uplifted spacetime $(\tilde{M},\tg_{AB})$,
\be
 \tilde{\Box}\Phi =
 0
 \,,
 \label{liftedLaplaceEq}
\ee
where $\tilde{\Box}\equiv \tg^{AB}\tilde{\nabla}_A\tilde{\nabla}_B$
and $\tilde{\nabla}_A$ are the d'Alembertian and Levi-Civita connection
 on $(\tilde{M},\tg_{AB})$.
Actually 
$\tilde{\Box}$ is calculated 
from (\ref{liftedHamiltonian})
as
\begin{equation}
\tilde{\square}_\tg
= \square_g
-2 qA^a \frac{\partial}{\partial x^a}\frac{\partial}{\partial v}
-q(\nabla_a A^a)\frac{\partial}{\partial v}
 +(q^2A_a A^a+2 V)\left( \frac{\partial}{\partial v} \right)^2
+2\frac{\partial}{\partial u}\frac{\partial}{\partial v} \,,
\end{equation}
where $\Box \equiv g^{ab}\nabla_a\nabla_b$ and $\nabla_a$
are the d'Alembertian and Levi-Civita connection on $(M,g_{ab})$.
Then restricting the solutions to the massless Klein-Gordon
equation (\ref{liftedLaplaceEq}) to the specific ones of the form
$
 \Phi =
 e^{-i\frac{m^2}{2}u+ iv}
 \psi({x^\rho})
$,
we have
\be
 \tilde{\Box}\Phi =
 e^{-i\frac{m^2}{2}u+ iv}
 \left({\mathcal H}+ m^2\right)
 \psi \,,
\ee
where ${\mathcal H}$ is the one appeared in (\ref{originalSchEq}).
For simplicity we set $m=0$ henceforth.
Thus, instead of solving the equation of motion (\ref{originalSchEq}),
we may solve the massless Klein-Gordon equation (\ref{liftedLaplaceEq}) on the uplifted spacetime $(\tilde{M},G_{AB})$.
This
fact
is the key in constructing symmetry operators of the equation of motion (\ref{originalSchEq}), as shown in the subsequent sections.

\subsection{Separability and commuting symmetry operators}
Generally, the separability of the differential equation
$ 
{\mathcal O}\psi = 0 
$
is related to the existence of first- and/or second-order differential
operators ${\mathcal K}^{(i)}$
satisfying the commutation relations
$
 [{\mathcal O},{\mathcal K}^{(i)}]  = 0 \,,
 [{\mathcal K}^{(i)},{\mathcal K}^{(j)}] = 0 \,,
$
where the square bracket $[\,\, , \,]$ denotes the commutator $[A,B]\equiv AB-BA$.
The operators satisfying the first condition are called symmetry operators since they map a solution of 
${\mathcal O}\psi = 0 $
to another solution, and the second condition implies that they commute with each other as differential operators.
If we are able to solve the differential equation ${\mathcal O}\psi = 0 $ by separation of variables, we have separation constants $\kappa$ and they are related to second-order symmetry operators by
$
 {\mathcal K}^{(i)}\psi = \kappa^{(i)} \psi \,.
$

\subsection{Construction of symmetry operators}
\label{sec:construct-sop}

In the Eisenhart-Duval lift, symmetry operators of the equation of motion (\ref{originalSchEq}) can be constructed from symmetry operators of the massless Klein-Gordon equation (\ref{liftedLaplaceEq}).
To show it, we begin with the second-order differential operator
 $\tilde{{\mathcal K}}$ on $\tilde{M}$,
\begin{equation}
 \tilde{{\mathcal K}} = \tilde{\nabla}_A \tilde{K}^{AB} \tilde{\nabla}_B \,,
 \label{symoponMtilde}
\end{equation}
where $\tilde{K}_{AB}$ is a rank-2 symmetric tensor on $\tilde{M}$.
For $\tilde{{\mathcal K}}$ to be a symmetry operator of the massless Klein-Gordon equation (\ref{liftedLaplaceEq}), i.e.,
\be
 [\tilde{\Box},\tilde{{\mathcal K}}] = 0 \,,
 \label{commutator_boxK}
\ee
$\tilde{K}_{AB}$ must be a Killing tensor
on $(\tilde{M},\tg_{AB})$, which obeys the Killing equation
$
\tilde{\nabla}_{(A}\tilde{K}_{BC)} = 0 \,,
$
and must satisfy, in addition, the anomaly-free condition
\be
 \tilde{\nabla}^A \left(\tilde{K}_{[A}{}^C\tilde{R}_{B]C}\right) = 0 \,,
 \label{anomalyfreecond}
\ee
where $\tilde{R}_{AB}$ is the Ricci tensor of $(\tilde{M},\tg_{AB})$~\cite{Carter:1977pq}.

Now, we suppose that
the Killing tensor
 $\tilde{K}^{AB}$
are independent of the coordinates $u$ and $v$,
$
 \partial_u \tilde{K}^{AB}=0 \,,
 \partial_v \tilde{K}^{AB}=0 \,.
$
Then, applying the symmetry operator $\tilde{{\mathcal K}}$ on $\tilde{M}$
to the wave function 
of the form 
$ \Phi =
 e^{-i\frac{m^2}{2}u+ iv}
 \psi({x^\rho})$,
we obtain the differential operator ${\mathcal K}$ on $M$ by
\begin{equation}
 \tilde{{\mathcal K}}\Phi = e^{iv} {\mathcal K}\psi \,.
    \label{symoponM}
\end{equation}
Without loss of generality, we may express $\tilde K^{AB}$ as
\footnotesize
\begin{align}
\nonumber
&
  (\tilde{K}^{AB})
=
\begin{pmatrix}
\tilde{K}^{ab} & \tilde{K}^{a u} & \tilde{K}^{a v}\\
\tilde{K}^{u b} & \tilde{K}^{uu} & \tilde{K}^{uv}\\
\tilde{K}^{vb} & \tilde{K}^{vu} & \tilde{K}^{vv}
\end{pmatrix}
 \\
&=
\begin{pmatrix}
K^{ab} & L^a &
-q K^{ac} A_c + M^a
 \\
L^b & W & -q L^c A_c + T + 2 WV
\\
-q K^{bc} A_c + M^b
&
-q L^c A_c + T + 2WV
&
q^2 K^{cd}A_c A_d - 2 q M^c A_c + F
\end{pmatrix}
.
\nonumber
\end{align}
\normalsize
Here,
using the fact that $\tilde K^{AB}$ is a Killing tensor on $\tilde M$,
we can show that
$K^{ab}$ and $L^a$ are Killing tensor and Killing vector on $M$, and also that $W$ is a constant and
$\nabla_a M^a = 0$ \cite{paper}.
Using $\nabla_a M^a = 0$ among them, we find
\begin{align}
{\mathcal K}\psi
&=
\nabla^{(A)}_a
K^{ab}
\nabla^{(A)}_b
\psi
+ 2i M^a 
\nabla^{(A)}_a
\psi
- F \psi
~,
\label{symop_explicit}
\end{align}
where $\nabla^{(A)}_a \equiv \nabla_a - iq A_a$.
Since 
$
 [\tilde{\Box}, \tilde{K}]\Phi = e^{iv} [{\mathcal H},{\mathcal K}]\psi \,,
$
we obtain
$
 [{\mathcal H},{\mathcal K}]=0 \,,
$
which means that ${\mathcal K}$ is a symmetry operator ${\mathcal K}$ of the equation of motion (\ref{originalSchEq}).

For commutativity of two symmetry operators 
$
\bigl[
\tilde{\mathcal K}^{(i)},
\tilde{\mathcal K}^{(j)}
\bigr]  =
0
$
additional conditions 
similar to (\ref{anomalyfreecond})
must be satisfied, which must be examined separately \cite{Kolar:2015cha}.

To summarize, the procedure for obtaining commuting symmetry operators ${\mathcal K}^{(i)}$ for the equation of motion (\ref{originalSchEq}) is given as follows:
\begin{enumerate}
 \item 
Read out $g_{ab}$, $A_a$ and $V$ from the equation of motion (\ref{originalSchEq}).
 \item Using these quantities,
  construct the uplifted metric $\tg_{AB}$ 
(\ref{liftedmetric}).
 \item Find Killing tensors $\tilde{K}^{(i)}_{AB}$ on ($\tilde{M},\tg_{AB}$)
 such that the components are independent of $u$ and $v$.
 \item 
 Check if $\tilde{K}^{(i)}_{AB}$ satisfy the anomaly-free condition (\ref{anomalyfreecond}),
and also the commutativity conditions.
 \item If they hold, construct the commuting symmetry operators $\tilde{{\mathcal K}}^{(i)}$
 by (\ref{symoponMtilde}).
 Then it follows that $[\tilde{\Box},\tilde{{\mathcal K}}^{(i)}]=0 = [\tilde{{\mathcal K}}^{(i)},\tilde{{\mathcal K}}^{(j)}]$.
 \item Finally, obtain commuting symmetry operators ${\mathcal K}^{(i)}$ from $\tilde{{\mathcal K}}^{(i)}$
 via (\ref{symoponM}) and (\ref{symop_explicit}). Then $[{\mathcal H},{\mathcal K}^{(i)}]=0=[{\mathcal K}^{(i)},{\mathcal K}^{(j)}]$.
\end{enumerate}

\section{Separability and symmetry operators of Maxwell's equations
 on the Kerr-NUT-(A)dS spacetime in four dimensions}
\label{sec:sec3}

To illustrate the procedure to construct the symmetry operators summarized in section~\ref{sec:construct-sop}, 
we apply it to the LKFK equation in four-dimensional case.
The Teukolsky equation in four dimensions and the LKFK equations can be treated in a similar manner.

\subsection{Carter form of the Kerr-NUT-(A)dS metric}
\label{sec:Carter4D}
The Kerr metric in the Boyer-Lindquist coordinates $(t,r,\theta,\phi)$ is 
\be
 ds^2 = 
-\frac{\Delta}{\Sigma}\left(dt - a\sin^2\theta d\phi\right)^2
 + \frac{\Sigma dr^2 }{\Delta}+ \Sigma d\theta^2
  + \frac{\sin^2\theta}{\Sigma}\left(adt - (r^2+a^2)d\phi\right)^2 \,,
\ee
where
$
 \Delta = r^2 -2mr+a^2 \,, 
 \Sigma = r^2 + a^2\cos^2\theta \,.
$
Performing the coordinate transformation
$
 p^2 = a^2\cos^2\theta \,, 
 \tau = t - a\phi
 \,, 
 \sigma = -\frac{\phi}{a}
 \,,
$
we obtain the Kerr metric in the Carter form
\begin{equation}
ds^2=-\frac{{\mathcal Q}(r)}{r^2+p^2}(d\tau-p^2 d\sigma)^2
+\frac{r^2+p^2}{{\mathcal Q}(r)}dr^2
 +\frac{r^2+p^2}{{\mathcal P}(p)}dp^2
+\frac{{\mathcal P}(p)}{r^2+p^2}(d\tau+r^2 d\sigma)^2 \,,
\label{Cartermetric}
\end{equation}
where 
$
 {\mathcal Q}(r) = r^2 -2 mr + a^2 \,, 
 {\mathcal P}(p) = a^2 - p^2
$.
The metric (\ref{Cartermetric}) solves the vacuum Einstein equation
with cosmological constant
$
 R_{ab}=\lambda g_{ab}
$
if and only if ${\mathcal Q}(r)$ and ${\mathcal P}(p)$ are given by
$
 {\mathcal Q}(r) = -\frac{\lambda}{3}r^4 + \epsilon r^2
 -2 mr + a^2 \,, 
 {\mathcal P}(p) = -\frac{\lambda}{3}p^4 - \epsilon p^2
 +2 lp + a^2 
$
with constants $\epsilon$, $a$, $m$ and $l$.
This solution is called the Kerr-NUT-(A)dS metric.
The metric (\ref{Cartermetric}) with ${\mathcal Q}(r)$ and ${\mathcal P}(p)$ replaced
by arbitrary functions depending only on $r$ and $p$ is called
the off-shell metric. In the following calculation,
we will work with the off-shell metric without using imposing the Einstein equation.

It is important to mention that the off-shell metric in the Carter
form fits into Benenti's canonical form for metrics
admitting the separability of the Hamilton-Jacobi equation for geodesics~\cite{MR1156532}.
Actually, the components of the inverse metric in the coordinate basis
are written in the canonical form 
\begin{equation}
\begin{aligned}
& g^{rr}= \frac{{\mathcal Q}}{r^2+p^2} \,, \quad
  g^{pp}= \frac{{\mathcal P}}{r^2+p^2} \,, 
\quad
g^{\tau \tau} = -g^{rr}\frac{r^4}{{\mathcal Q}^2}
  +g^{pp}\frac{p^4}{{\mathcal P}^2} \,, 
\\
&  g^{\tau \sigma} = g^{rr}\frac{r^2}{{\mathcal Q}^2}
  +g^{pp}\frac{p^2}{{\mathcal P}^2} \,, \quad
  g^{\sigma \sigma} = -g^{rr}\frac{1}{{\mathcal Q}^2}
  +g^{pp}\frac{1}{{\mathcal P}^2} \,,
\end{aligned}
\label{g-Carter4D}
\end{equation}
and we can easily construct a Killing tensor
on the Kerr spacetime.
Following Benenti's construction,
the Killing tensor
turns out to be given by (\ref{g-Carter4D}) with 
$g^{ab}, g^{rr}, g^{pp}$ 
replaced to 
$K^{ab}, K^{rr}, K^{pp}$, where 
\begin{equation}
 K^{rr} = \frac{p^2 {\mathcal Q}}{r^2+p^2} \,, \quad
  K^{pp} = -\frac{r^2 {\mathcal P}}{r^2+p^2}~.
\end{equation}
%


\subsection{Symmetry operators of the 
LKFK
equations}

In the LKFK formulation in four dimensions,
Maxwell's equation (\ref{Maxwell}) can be written in the form
\be
 {\mathcal H}^{(L)} \psi
 \equiv \Big[g^{ab}(\nabla_a - i qA^{(L)}_a)
 (\nabla_b - i qA^{(L)}_b)
 - 2V^{(L)}\Big]\psi = 0 \,,
\label{Meq-gauged}
\ee
where $g_{ab}$ is the off-shell metric (\ref{Cartermetric}),
and $A^{(L)}$ and $V^{(L)}$ are given by
\footnotesize
\begin{equation}
\begin{aligned}
 qA^{(L)}_a dx^a =&
 \frac{i\beta^2 r}{\Xi_r}dr-\frac{i\beta^2p}{\Xi_p}dp
 - \frac{i\beta}{r^2+p^2}\left(\frac{{\mathcal Q}}{\Xi_r}
 -\frac{{\mathcal P}}{\Xi_p}\right)d\tau
 +\frac{i\beta}{r^2+p^2}\left(\frac{p^2{\mathcal Q}}{\Xi_r}
 +\frac{r^2{\mathcal P}}{\Xi_p}\right)d\sigma \,, \\
 V^{(L)} =& \frac{1}{r^2+p^2}\Bigg(
 \frac{\beta^2 r{\mathcal Q}'}{2\Xi_r}
 +\frac{\beta^4 r^2{\mathcal Q}}{\Xi_r^2}
 -\frac{\beta^2 p {\mathcal P}'}{2\Xi_p}
 +\frac{\beta^4 p^2{\mathcal P}}{\Xi_p^2}\Bigg) \,,
\end{aligned}
\label{ALVL}
\end{equation}
\normalsize
where
$
 \Xi_r = 1-\beta^2r^2 \,, 
 \Xi_p = 1+\beta^2p^2 
$.
This can be solved by separation of variables by using
the ansatz 
$ \psi = e^{-i\omega \tau}e^{im\sigma} R(r)\Theta(p)$
and we obtain
\begin{equation}
\begin{aligned}
\Xi_r
  {\mathcal Q}\frac{d }{dr}
  \left(
  \frac{\mathcal Q}{\Xi_r}\frac{dR}{dr}
  \right)
  +
\left[
(\omega r^2 + m)^2
  - \frac{2i{\mathcal Q}}{\beta\Xi_r}(\omega +\beta^2  m)
  - \kappa^{(L)}{\mathcal Q}
  \right] R
&  = 0 \,,
\\
 \Xi_p
   {\mathcal P}\frac{d }{dp}
   \left(
   \frac{\mathcal P}{\Xi_p}\frac{d\Theta}{dp}
   \right)
 + \left[
 - ( \omega p^2 - m)^2
 + \frac{2i{\mathcal P}}{\beta\Xi_p}(\omega + \beta^2  m)
+ \kappa^{(L)}{\mathcal P}
\right]\Theta
&  = 0 \,,
\end{aligned} 
\label{LKFKeq_explicit}
\end{equation}
where $\kappa^{(L)}$ is a separation constant.


The uplifted metric $\tg_{AB}$ on $\tilde{M}$ is obtained from (\ref{liftedmetric}) with (\ref{Cartermetric}) and (\ref{ALVL}). As explained before, the uplifted metric $\tg_{AB}$ fits into Benenti's canonical form
\footnotesize
\begin{align}
\label{G_LKFK4D}
& \tg^{ab} = g^{ab} \,, \qquad
  \tg^{rv} = -g^{rr} \frac{i \beta^2 r}{\Xi_r} \,, \qquad
  \tg^{pv} = g^{pp} \frac{i \beta^2 p}{\Xi_p} \,, \\
& \tg^{\tau v} = - g^{rr}\frac{i}{2 \beta} \frac{1+\beta^2 r^2}{\Xi_r {\mathcal Q}}
  + g^{pp}\frac{i}{2 \beta} \frac{1-\beta^2 p^2}{\Xi_p {\mathcal P}} \,, \qquad
  \tg^{\sigma v} = g^{rr}\frac{i\beta}{2} \frac{1+\beta^2 r^2}{\Xi_r {\mathcal Q}}
  - g^{pp}\frac{i \beta}{2} \frac{1-\beta^2 p^2}{\Xi_p {\mathcal P}} \,, \nonumber\\
& \tg^{uv} = 1 = g^{rr}\frac{r^2}{{\mathcal Q}}+g^{pp} \frac{p^2}{{\mathcal P}} \,, \qquad
  \tg^{vv} =
  \frac{g^{rr}}{{\mathcal Q}}
  \frac{d}{dr}
\left( \frac{\beta^2 r}{\Xi_r}  {\mathcal Q}\right)
  -
  \frac{g^{pp}}{{\mathcal P}}
  \frac{d}{dp}
  \left( \frac{\beta^2 p}{\Xi_p} {\mathcal P} \right)
   \,,
\nonumber
\end{align}
\normalsize
and the Killing tensor is given by the expression same as the metric (\ref{G_LKFK4D}) with 
$\tilde g^{AB}, g^{ab}, g^{rr}, g^{pp}$ replaced to
$\tilde K^{AB}, K^{ab}, K^{rr}, K^{pp}$, respectively.
Then we obtain the symmetry operator of 
the LKFK equations
according to (\ref{symop_explicit}) as
\footnotesize
\begin{align}
 {\mathcal K}^{(L)}\psi
 =& \frac{p^2}{r^2+p^2}\Bigg[
 {\mathcal Q}\partial_r^2
 + \left({\mathcal Q}'+\frac{2\beta^2 r {\mathcal Q}}{\Xi_r}\right)\partial_r
 -\frac{1}{{\mathcal Q}}(r^2\partial_\tau - \partial_\sigma)^2
 + \frac{2}{\beta}\frac{1}{\Xi_r}(\partial_\tau-\beta^2\partial_\sigma)
 \Bigg]\psi
 \\
 & -\frac{r^2}{r^2+p^2}\Bigg[
 {\mathcal P}\partial_p^2
 + \left({\mathcal P}'-\frac{2\beta^2 p {\mathcal P}}{\Xi_p}\right)\partial_p
 + \frac{1}{{\mathcal P}}(p^2\partial_\tau + \partial_\sigma)^2
 -\frac{2}{\beta}\frac{1}{\Xi_p}(\partial_\tau-\beta^2\partial_\sigma)
 \Bigg]\psi \,.
\nonumber
\end{align}
\normalsize
Using the multiplicative separation ansatz $ \psi = e^{-i\omega \tau}e^{im\sigma} R(r)\Theta(p)$
and the separated equations (\ref{LKFKeq_explicit}),
we find
\be
 {\mathcal K}^{(L)}\psi = \kappa^{(L)} \psi \,.
\ee
It can be confirmed by direct calculations that the anomaly-free condition (\ref{anomalyfreecond}) 
is satisfied by the Killing tensor constructed above \cite{paper},
hence
we have the commutation relations
\be
 [{\mathcal H}^{(L)},{\mathcal K}^{(L)}] = 0 \,.
\ee

\section{Outlook}

An obvious next step 
for this work
is to inspect the uplifted metrics corresponding to different perturbations.
In four dimensions Maxwell's equations can be reduced to the Teukolsky equation and also to the LKFK equations, and these two equations lead to different uplifted metrics.
Since these two equations should describe the same time evolution of the Maxwell fields, 
there would be intertwining operators connecting the solutions of these solutions. 
If they exist, it is expected that the uplifted metrics corresponding to the different equations of motion are related to each other by some transformations.
It would be fruitful to investigate such possible connection between the uplifted spacetimes by clarifying their geometric properties such as the curvature invariants and Weyl tensors.

Also, for both equations the uplifted metric is intimately related to the symmetry of the base spacetime. For example, the vector field for the LKFK equation is given by $A_a \sim \xi^b B_{ba}$, where $\xi^b$ is a Killing vector and the polarization tensor $B_{ab}$ is constructed from the metric and the principal tensor corresponding to the hidden symmetry of the base spacetime. Geometric properties of this kind could provide useful information of properties of the uplifted metrics and also the vector potential ansatz that reduced the Maxwell equation to a scalar field equation. Pursuing this idea would be another future tasks for this study.

\end{document}